\documentclass[apj]{openjournal}

\usepackage{xcolor}
\usepackage{textgreek}
\usepackage[utf8]{inputenc}
\usepackage[english]{babel}

\usepackage{hyperref,amsmath}
\hypersetup{
    unicode, 
    colorlinks=true,
    linkcolor=linkcolor,
    citecolor=linkcolor,
    filecolor=linkcolor,
    urlcolor=linkcolor,
}
\usepackage{color,colortbl}
\definecolor{linkcolor}{rgb}{0.0,0.3,0.5}
\usepackage{tensind}
\tensordelimiter{?}
\DeclareGraphicsExtensions{.bmp,.png,.jpg,.pdf}
\usepackage{verbatim}
\usepackage[normalem]{ulem}
\usepackage{orcidlink}
\usepackage{soul}

\graphicspath{ {./figs/} }

\begin{document}
\title{The $\tau$ of Neutral Hydrogen: Increased CMB Optical Depth at Long Wavelengths}

\author{Gilbert Holder \orcidlink{0000-0002-0463-6394}}
\affiliation{Department of Physics, University of Illinois Urbana-Champaign, 1110 W Green St, Urbana, IL 61821, USA}

\author{Adrian Liu \orcidlink{0000-0001-6876-0928}}
\affiliation{Department of Physics and Trottier Space Institute, McGill University, 3600 Rue University, Montreal, QC H3A 2T8, Canada}

\author{Tzu-Ching Chang \orcidlink{0000-0001-5929-4187}}
\affiliation{Jet Propulsion Laboratory, California Institute of Technology, 4800 Oak Grove Dr., Pasadena, CA 91109, USA}
\affiliation{Department of Physics, California Institute of Technology, 1200 E. California Boulevard, Pasadena, CA 91125, USA}

\author{David Zegeye \orcidlink{0000-0002-4263-340X}}
\affiliation{Department of Physics, University of Illinois Urbana-Champaign, 1110 W Green St, Urbana, IL 61821, USA}

\begin{abstract}
    At wavelengths longer than $21\,{\rm cm}$, photons from the long-wavelength tail of the cosmic microwave background (CMB) have a non-zero probability of being absorbed by distant neutral hydrogen. This provides an additional suppression of the observed CMB clustering in addition to the usual Thomson scattering. The optical depth as a function of frequency is sensitive to the $21\,{\rm cm}$ spin temperature $T_s$ of the gas as a function of cosmic time, with the excess optical depth peaking at a level of a few percent around 100 MHz. The details depend on the specifics of the heating of cosmic gas and the evolution of the neutral fraction $x_{\rm HI}$. It is likely difficult to detect the CMB at these long radio wavelengths, but the cause is aided by the ability to cross-correlate with the already well-characterized fluctuations at cm/mm frequencies. We find that detecting CMB fluctuations at radio wavelengths corresponding to the $21\,{\rm cm}$ ``dark ages'' in cross-correlation with mm-wave maps may be easier than detecting the intrinsic 21 cm fluctuations. Measurement of the amplitude of CMB fluctuations as a function of radio wavelength provides a path for a new type of direct measurement of the combination $x_{\rm HI}/T_s$ as a function of redshift.
\end{abstract}


\maketitle

\section{Introduction}
\label{sec:intro}

The cosmic microwave background (CMB) is the remnant radiation from the early universe, now redshifted to peak at mm-wavelengths \citep{2002ARA&A..40..171H}. The observed fluctuations have been a treasure trove of information about the early universe and up to the time of ``last scattering,'' when electrons became locked up with protons to form hydrogen atoms. From those early times to today, CMB photons have largely freely propagated across the universe, being gently deflected by large scale gravitational potentials and occasionally scattering with neutral atoms or remaining free electrons. At relatively late times, the first stars are expected to have produced enough ionizing photons to once again produce a mainly ionized universe, through a process called reionization \citep{2006ARA&A..44..415F,2013fgu..book.....L}. The Thomson scattering probability in the reionized universe is sufficiently high that a noticeable fraction of CMB photons are thought to have scattered. This produces a suppression of the observed CMB angular power spectrum, as fluctuations behind the electron screen are scattered out of the line of sight while the photons scattered into our beam will not have these fluctuations.

The Thomson scattering probability is parameterized by the optical depth, $\tau_e$, and is an important parameter for cosmology, as it allows us to relate the observed CMB fluctuations to other measures of the fluctuations in gravitational potentials. Thomson scattering also leads to large-angle CMB polarization fluctuations, which have been used to measure $\tau_e$ \citep{2020A&A...641A...6P} to be around 0.06.

At longer wavelengths, there is additional opacity coming from resonant $21\,{\rm cm}$ absorption by neutral hydrogen. The absorption probability depends on the relative level populations, characterized by the $21\,{\rm cm}$ spin temperature $T_s$. The evolution of $T_s$ over cosmic time is a bit complicated (see \citealt{2006PhR...433..181F,2006MNRAS.367..259H,2010ARA&A..48..127M,2012RPPh...75h6901P,2013fgu..book.....L}). At early times (but after recombination), the spin temperature is expected to be coupled to the gas temperature, while the gas temperature tracks the CMB temperature through residual photon-electron scatterings on the small residual electron population. At later times, the gas density and electron fraction are sufficiently low that this process becomes insufficient to keep the gas coupled to the CMB, and the gas cools adiabatically, becoming colder than the CMB. As the density drops, collisions become too rare to keep the spin temperature coupled to the gas temperature, at which point the spin temperature comes into radiative equilibrium with the CMB. Subsequently, stars start to form, producing all sorts of new electrons and photons, and the spin temperature can fluctuate depending on the physics of reionization. This physics is sufficiently rich that measurement of this signal is the goal of several ongoing experiments aiming to measure this signal by tracking the monopole of the intensity of the radio sky as a function of wavelength \citep{2018MNRAS.478.4193P,2019JAI.....850004P,2021RSPTA.37990566C,2022NatAs...6..984D,2022NatAs...6..607S,2023ExA....56..741S,2024MNRAS.530.4125M,2025RASTI...4af046B,2025arXiv250911846M,2025PASP..137l5002C,2025RASTI...4...61A,2026arXiv260202661B}.

A low spin temperature leads to a high probability of absorption and therefore a relatively high optical depth, which we will call $\tau_{\rm HI}$. Just like the case of electron scattering, this $21\,{\rm cm}$ absorption will suppress the observed primary CMB fluctuations, with the observed power spectrum being lower than the power spectrum measured at mm-wavelengths by a factor $e^{-2\tau_{\rm HI}}$. A cross-power spectrum with mm-wave CMB measurements would yield a power spectrum that is reduced by a factor of $e^{-\tau_{\rm HI}}$ as compared to the mm-wave power spectrum. 

In the rest of this paper, we will calculate the expected size of this effect and give some rough estimates of how this signal compares to other signals and noise.
These results are not strongly dependent on cosmological parameters; for calculations that follow we assume cosmological parameters $\Omega_b h^2=0.02237$,$\Omega_c h^2=0.12$, $h=0.6736$, $n_s=0.9649$, $A_s=2.1\times 10^{-9}$, $\tau_e=0.0544$, and $Y_p=0.2454$ \citep{2020A&A...641A...6P}.

\section{Neutral Hydrogen Optical Depth}
\label{sec:tau21}

The $21\,{\rm cm}$ hyperfine transition occurs at a rest frequency of $\nu_\circ=1420$\,MHz. Photons that we observe today at lower frequencies had some time in the past when they were at or close to a frequency of $\nu_\circ$, when they would have had the chance to be absorbed by neutral hydrogen in the hyperfine ground state. Lower $T_s$ leads to an increased ground state fraction and thus increased absorption probability, as does a higher $x_{\rm HI}$, the fraction of hydrogen that is neutral (as opposed to ionized). The absorption we consider in this paper is that from gas at cosmological mean density. There will be fluctuations in this absorption probability coming from density and velocity fluctuations, but for this paper we will ignore these perturbations.\footnote{In essence, we are considering the effect of a uniform background of gas on an anistropic CMB background. This is the opposite limit to typical $21\,\textrm{cm}$ cosmology measurements, which are targeting small-scale fluctuations in gas density, velocity, and ionization imprinted on a uniform CMB background.} The probability of photons at observed frequency $\nu$  below $\nu_\circ$ having been observed by cosmological gas in Hubble flow at the cosmic mean at redshift $z$ is given by \citep{2006PhR...433..181F}
\begin{equation}
    \tau_{\rm HI}\sim 0.00847 (1+z)^{3 \over 2}{x_{\rm HI} \over T_s} 
    \Bigl({\Omega_b h^2 \over 0.2237}\Bigr)
    \Bigl({1-Y_p \over 0.7546}\Bigr)
    \sqrt{{0.1424 \over \Omega_c h^2+\Omega_b h^2}} \ ,
    \label{eqn:tau}
\end{equation}
where we have used slightly different fiducial cosmological parameters from those of \citet{2006PhR...433..181F}.
\begin{center}
    \begin{figure}[!t]
        \centering
    	\includegraphics{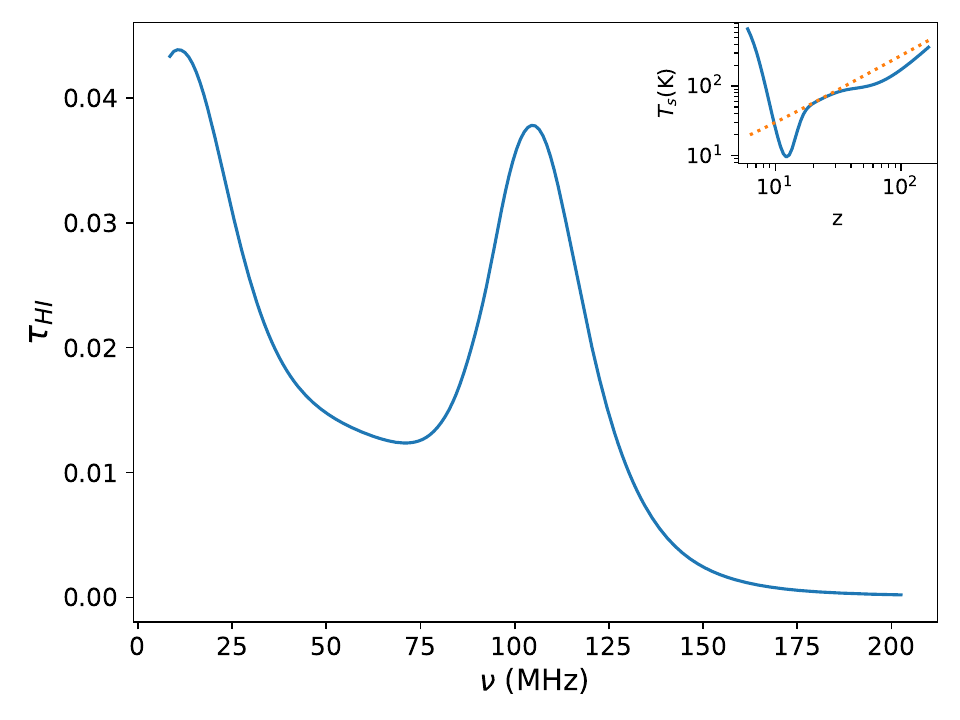}
    	\caption{Optical depth to neutral hydrogen as a function of redshift for the evolution of the spin temperature shown in the inset. In the inset, the CMB temperature as a function of redshift is shown for reference as the dotted power-law.}
        \label{fig:tau}
    \end{figure}
\end{center}

The evolution of $T_s$ with redshift depends on several physical effects, determined to be a combination of the CMB temperature $T_{\rm CMB}(z)$ and the gas temperature $T_K(z)$. Shortly after recombination, $T_K=T_{\rm CMB}$ until $z \lesssim 200$, when the residual ionization drops sufficiently low that there aren't enough electrons to maintain coupling between the gas and the CMB. From that point, the gas cools faster than the CMB, by adiabatic expansion, and collisions between neutral hydrogen atoms are sufficiently rapid for $T_s$ to track $T_K$. Stars do not form until lower redshifts, at $z \sim 65$ for \emph{the} first star in our Hubble volume \citep{2006MNRAS.373L..98N}, and saturation of sustained formation in first-generation Population III stars at $20 \lesssim z \lesssim 30$ \citep{2009ApJ...694..879T}. Thus, until then there is an extended period when $T_s<T_{\rm CMB}$ and neutral hydrogen can in principle be seen in absorption relative to the CMB, both in the mean spectrum at low frequencies and as spatial fluctuations tracing out the density inhomogeneities. This period is known as the ``dark ages'' to emphasize that to this point there have been no stars \citep{2007PhRvD..76h3005L,2018PhRvD..98d3520B,2024RSPTA.38230068F}.

At $z<50$ the collisions between gas particles start to become rare as the density continues to drop, and eventually the most frequent interaction for neutral hydrogen atoms are with CMB photons, so $T_s\sim T_{\rm CMB}$, and the gas again becomes difficult to detect against the CMB. 

Eventually, stars will start to form, creating new photons in the universe, including X-rays that can heat the cosmic gas and UV photons that can lead to atomic transitions and ionizations of neutral hydrogen atoms. This period is known as ``cosmic dawn.'' At this point, things get complicated, but it is plausible that the UV interactions will again couple $T_s$ to $T_K$, which will initially still be relatively cold (much colder than the CMB) if gas heating has only just begun. The neutral hydrogen will again be visible in absorption relative to the CMB, but as the gas starts to heat up the temperature will eventually exceed that of the CMB, at which point the neutral hydrogen will be visible in emission compared to the CMB.\footnote{There exist \emph{cold reionization} models where the intergalactic medium is negligibly heated \citep{2017ApJ...840...39M,2017MNRAS.464.1365M}, and these would lead to a larger absorption dip and therefore a larger amplitude for the signal discussed in this paper. However, recent upper limits on the $21\,\textrm{cm}$ power spectrum have ruled out strictly cold scenarios \citep{2020MNRAS.493.4728G,2021MNRAS.503.4551G,2021MNRAS.500.5322G,2021MNRAS.501....1G,2022ApJ...925..221A,2023ApJ...945..124H,2025A&A...699A.109G,2025ApJ...989...57N,2026ApJ...998...33A} although some parameter space remains for models where heating is relatively mild.} Eventually, the universe will have produced enough UV photons that the neutral hydrogen will be largely cleared, the time known as the ``end of reionization.''

The net effect of all of this is that the spin temperature will track $T_{\rm CMB}$ until dipping below it for a while, returning back to $T_{\rm CMB}$ again, then again dipping strongly below the CMB briefly before spiking up to high temperatures. This behavior is seen in the inset of Figure \ref{fig:tau} for a possible scenario. The ``dark ages'' part ($z\sim 40$ to $150$) is set by well-determined physics and is relatively robust, but the details of ``cosmic dawn'' can greatly affect the evolution of $T_s$.

The optical depth $\tau_{\rm HI}$ is shown in 
(Fig.~\ref{fig:tau}) as a function of frequency for the particular $T_s$ evolution shown in the inset. The $T_s$ dip at high redshifts corresponds to the lowest frequencies shown, while the dip in $T_s$ at $z\sim 10-20$ can be seen as the peak around 100 MHz. From the form of Equation \ref{eqn:tau} we can see that lower $T_s$ leads to higher $\tau_{\rm HI}$, so the detailed shape of $\tau_{\rm HI}(\nu)$ will depend sensitively on how effectively the UV photons can couple $T_s$ and $T_K$ while the gas also begins to be heated and eventually ionized. As can be seen from Equation \ref{eqn:tau}, measuring $\tau_{\rm HI}(z)$ provides a direct measurement of the quantity $x_{\rm HI}/T_s$. In our calculations we assume a neutral universe, as we are primarily focused on the times before reionization is well underway. 

This additional optical depth means that the low-frequency CMB fluctuations will be slightly suppressed.  Photons coming from a hot or cold spot in the CMB will have a slightly boosted probability of being absorbed at low frequencies as compared to at mm-wavelengths. Any re-emission would be done nearly isotropically, so the observed angular fluctuations in the intensity at these frequencies will be slightly reduced. The CMB power spectrum at radio wavelengths will be suppressed by a factor of $e^{2\tau_{HI}}$ while the cross power between radio and mm-wave measurements will be reduced by $e^{\tau_{HI}}$. As can be seen in Figure \ref{fig:tau} this reduction in cross power will be at the level of 3.5\% for radio observations at the frequencies corresponding to cosmic dawn and can be above 4\% for dark ages experiments. 

This optical depth at long wavelengths is in addition to the usual late-universe scattering by free electrons which is frequency-independent and of great concern to cosmologists using the mm-wave CMB fluctuations. Unlike the case of Thomson scattering, there is no accompanying additional large-angle polarization from $\tau_{\rm HI}$.

\section{Detectability}

We have shown that the CMB fluctuations will be a few percent dimmer at long wavelengths, but this is obviously irrelevant if the CMB fluctuations are unmeasurable at these wavelengths. 

The CMB was first measured at radio frequencies around 4 GHz \citep{1965ApJ...142..419P}, but it was quickly realized that higher frequencies were better suited for these measurements: the CMB spectrum peaks at mm-wavelengths while foregrounds from synchrotron and free-free emission fall steeply with increasing frequency compared to the CMB. For example, the \emph{Planck} satellite, which provided the precise measurements that allowed the determination of the cosmological parameters that we use in this paper \citep{2020A&A...641A...6P}, performed measurements of CMB anisotropies using channels that ranged from  frequencies of 30 GHz to 857 GHz, with the most constraining frequencies being the channels around 100, 143, and 217 GHz. 

Is it possible to measure the CMB fluctuations at frequencies three orders of magnitude lower than the typical frequencies used for precise CMB measurements? 
Intensity fluctuations from Galactic and extragalactic synchrotron radiation will be many orders of magnitude brighter than the CMB fluctuations at these wavelengths. As we will argue below, the answer is not a ``hard no,'' but there are certainly some things to worry about. 

The only thing that makes this even remotely possible is that the CMB fluctuations are measured extremely well at mm-wavelengths already. We have a very high signal-to-noise template for what we expect the CMB to look like at radio wavelengths, with the only free parameter being a small uncertainty in the amplitude that comes from this extra optical depth that slightly suppresses the radio CMB fluctuations. Importantly, this suppression of power affects all scales equally. This template effectively allows us to do a weighted average of all the pixels in our radio map, squeezing all of the map information into a measurement of a single template amplitude. For measurements with a lot of per-pixel noise, this is a huge improvement over the requirements for trying to measure the power spectrum using only the noisy data.  
Switching to a harmonic space description, in the limit where the noise power $N_\ell$ per mode is much larger than the signal power $S_\ell$, the signal-to-noise on the amplitude scales as $\sqrt{S_\ell/N_\ell}$ per mode in cross-correlation vs $S_\ell/N_\ell$ using the auto-correlation.

The effective noise in the radio map in cross-correlation is similar to the per-pixel noise divided by the square root of the number of effective pixels in the radio map. The sub-degree CMB intensity fluctuations have root mean square (rms) fluctuations around 100 $\mu$K, so measurements over several thousand square degrees with an rms of 10s of mK would be sufficient to detect the CMB in cross-correlation with a mm-wave CMB template. To detect the CMB polarization fluctuations in cross-correlation would require roughly an order of magnitude deeper maps.

Simply detecting the CMB fluctuations at long radio wavelengths would be a remarkable feat, but the real science lies deeper, as we want to detect the small differences in cross-correlation amplitude as a function of frequency to measure the variation in $\tau_{\rm HI}(\nu)$. Roughly speaking, this means that we would want to detect this at tens to a hundred ``$\sigma$'' to be able to precisely measure $x_{\rm HI}/T_s$ as a function of redshift. 

Somewhat paradoxically, it may be easier to measure small changes in $\tau_{\rm HI}$ than it would be to measure the overall CMB fluctuations, as the changes in $\tau_{\rm HI}$ will lead to variations in each pixel in the map as a function of frequency. (See, for example, the proposal in \citealt{2025PhRvD.111f3517Z} to measure CMB spectral distortions with low-frequency radio measurements). The frequency dependence of $\tau_{\rm HI}$ is helpful, as the radio sky is very bright in synchrotron emission, on the order of 1000\,K at these frequencies \citep{2008MNRAS.388..247D, 2017MNRAS.464.3486Z,2017MNRAS.469.4537D}. This causes two problems: increased load on the detectors which leads to more noise, and more problematically the spatial variations in the synchrotron emission will lead to variations of many Kelvin from pixel to pixel from varying amounts of emission. The leading idea for removing these variations is to take advantage of the spectral smoothness of this foreground: while it is very bright, it is smoothly varying across a broad range in frequencies and could in principle be greatly reduced by focusing only on signals that vary noticeably between closely spaced frequencies (see \citealt{2020PASP..132f2001L} for a review of a broad class of foreground mitigation strategies). 

Foreground removal for low-frequency radio cosmology experiments is an active area of research. Most efforts to date focus on searching for the fluctuations induced by the hydrogen fluctuations to be detected in the auto-correlation. It is possible that different techniques will be needed for the case where a good template for the signal is available to be used for cross-correlation. We leave the hard work of realistic foreground removal concerns to future work (possibly by other authors) and instead do a toy comparison of the difficulty of detecting the CMB in cross-correlation as compared to detecting the $21\,{\rm cm}$ intrinsic fluctuations at similar wavelengths. We will consider two cases, $z=50$ (dark ages) and $z=15$ (cosmic dawn) and take a single channel with a fractional bandwidth of 1\%. This is a somewhat broad bandwidth for a $21\,{\rm cm}$ experiment and a narrow one for a CMB experiment, but the motivation is that perhaps other adjacent channels could be used to remove foreground emission over this 1\% band without greatly reducing the CMB sensitivity.

\begin{figure}
\centering{\includegraphics{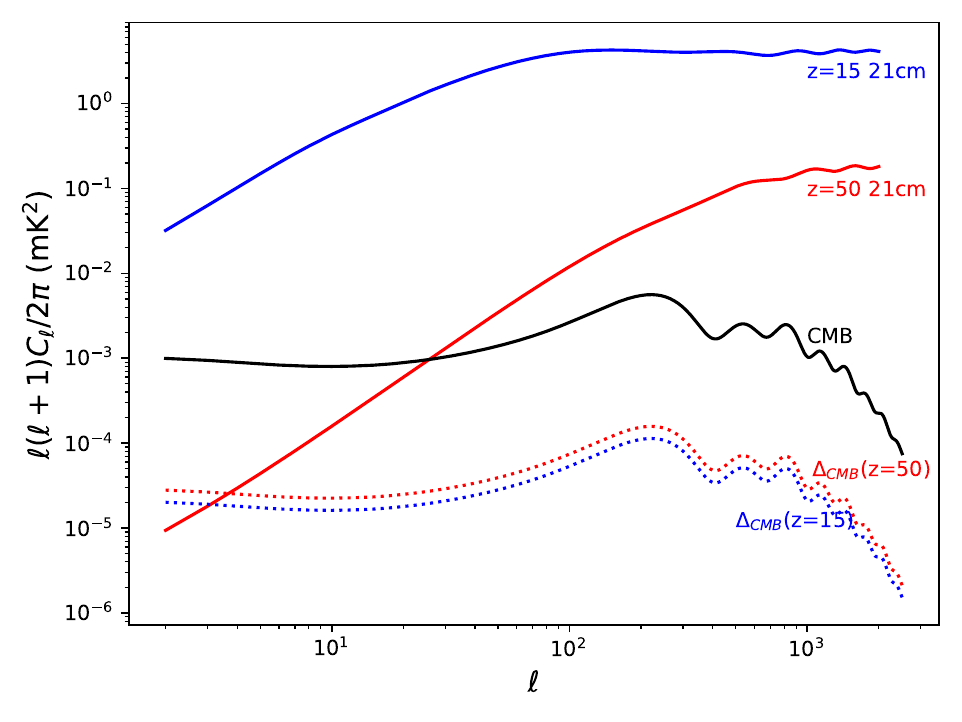}}
    \caption{Angular power spectra for the $21\,{\rm cm}$ autospectra at z=50 ($\nu\sim28$ MHz, red) and z=15 ($\nu\sim 89$ MHz, blue), compared to the CMB fluctuations (in the middle of the plot) and the differential CMB fluctuations at frequencies corresponding to $21\,{\rm cm}$ fluctuations at z=50 and z=15. For the $21\,{\rm cm}$ power spectra a fractional bandwidth of 1\% was assumed. }
    \label{fig:cl}
\end{figure}

To compare the detectability of the $21\,{\rm cm}$ auto-correlation and the radio CMB, we need to calculate the expected level of the $21\,{\rm cm}$ fluctuations. To do this, we use CAMB to calculate the angular power spectrum of the fractional $21\,{\rm cm}$ fluctuations \citep{2007PhRvD..76h3005L} $\Delta T/T$ at the redshifts of interest and multiply the derived angular power spectrum by the square of the expected brightness temperature, given by \citep{2006PhR...433..181F}
\begin{equation}
    T_b(z) = {\tau_{\rm HI} \over 1+z} (T_s-T_{\rm CMB}) \ .
\end{equation}
The angular power spectra are shown in Figure \ref{fig:cl}. The power at $z=15$ for this model is well above the CMB fluctuations, such that the $21\,{\rm cm}$ fluctuations would need to be imaged at high signal-to-noise in order for the CMB fluctuations to be detectable, at which point the $21\,{\rm cm}$ fluctuations would be a substantial source of sample variance in the cross-correlation. 

As a very simple estimate of signal-to-noise, we assume a fractional bandwidth of 1\% and basic white noise ($C_\ell$ being constant), expressed in units of mK-arcmin. This unit expresses the rms temperature noise in a pixel of area equal to one square arcminute. Subdividing this pixel into 4 smaller pixels would lead to pixels of 1/4 the area and double the rms temperature noise per new pixel, while binning 4 adjacent pixels into a pixel of 4 times the area would give half the rms temperature noise. We ignore beam effects for this calculation, including all modes up to $\ell=2000$. The maximum $\ell$ has only a small effect, due to the $\ell$-shape of the signals. The only exception is in the very high signal-to-noise limit of the case of detecting the $21\,{\rm cm}$ auto-correlation. 

To calculate the expected signal-to-noise for auto-clustering, we use \citep{1995PhRvD..52.4307K} the expected signal-to-noise per $\ell$:
\begin{equation}
    \Bigl({S \over N} \Bigr)^2 = \sum_\ell {(2\ell+1) f_{sky} \over 2} {S_\ell^2 \over (S_\ell+N_\ell)^2} \ .  
\end{equation}

For the signal-to-noise in the cross power for two maps $X$ and $Y$ the signal-to-noise is given by
\begin{equation}
    \Bigl({S \over N} \Bigr)^2 = \sum_\ell {(2\ell+1) f_{sky} 
    }
    {S^{(XY)\, 2}_\ell \over S^{(XY)\, 2}_\ell+(S^X_\ell+N^X_\ell)(S^Y_\ell+N^Y_\ell)} \ ,  
\end{equation}
where the superscripts indicate signal and noise power spectra for each map and the cross power spectrum between the maps is $S^{(XY)}_\ell$. In the particular case at hand, we assume that our CMB map is taken from mm-wave observations and has no noise for the purposes of this cross-correlation. Furthermore, for the case of cross-correlation of the CMB signal in the two maps, $S^{XY}_\ell$ is simply the CMB power spectrum as measured by Planck ever so slightly attenuated by the excess optical depth $e^{-\tau_{\rm HI}(\nu)}$.

\begin{figure}
\centering{\includegraphics{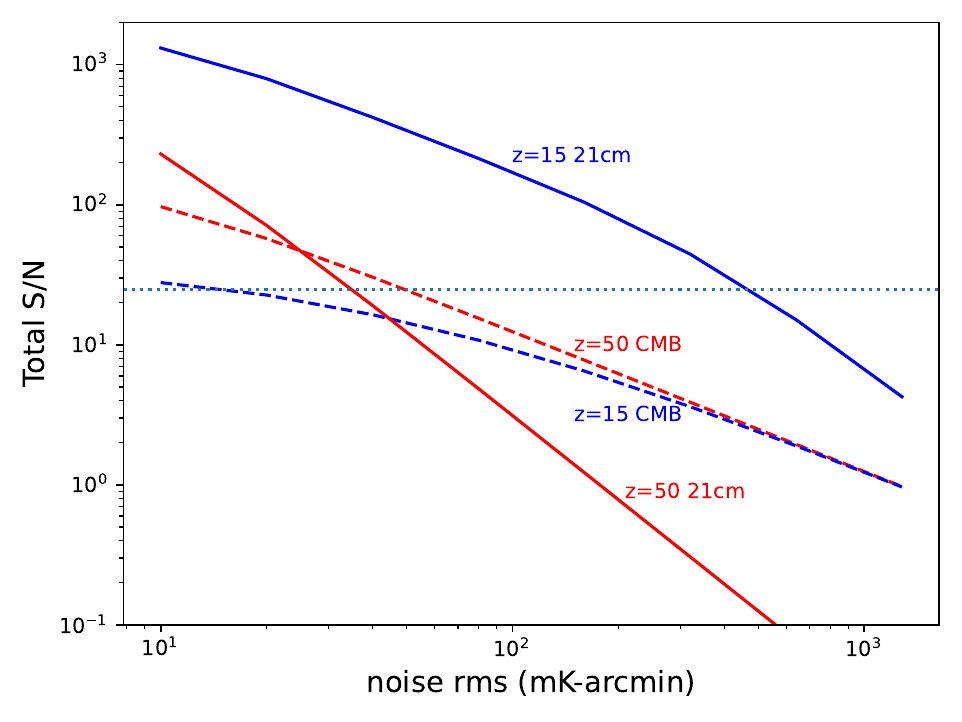}}
    \caption{Total signal-to-noise as a function of map noise for detecting either the CMB in cross-correlation (dashed curves) or the auto-correlation of the $21\,{\rm cm}$ signal, assuming a full sky measurement. The dotted line shows a nominal threshold of 25 for detecting the CMB additional optical depth $\tau_{\rm HI}$ in cross-correlation. The x-axis is the rms map noise in units expressing the noise in mK in a 1'x1' pixel. For the 21 cm signal we assume a 1\% bandwidth.}
    \label{fig:snr}
\end{figure}

To achieve a signal-to-noise of 25 in the CMB cross-correlation at $\nu\sim 89$ MHz, the signal-to-noise in the $21\,{\rm cm}$ auto-correlation would be well over 1000, as seen in Figure \ref{fig:snr} when the dashed blue curve rolls over at low noise levels.

In principle, it would be possible to use polarization measurements for measuring the dark ages CMB optical depth through the E-mode polarization of the CMB. By the time this is possible there will surely exist nearly perfect E-mode polarization templates at mm-wavelengths. This would allow the EE cross-correlation to be used (and to a lesser extent the radio E modes correlated with CMB intensity fluctuations using the TE correlation).  However, the CMB is roughly 10\% polarized, knocking the expected signal to noise down by an order of magnitude for the polarized CMB. At very low noise levels (at the level of a few mK-arcmin), the gain in beating sample variance could potentially lead to a comparable detection using Stokes Q/U instead of Stokes I fluctuations. 

For the dark ages signal, the $21\,{\rm cm}$ power spectrum is somewhat above the CMB power spectrum, but the advantages of cross-correlation vs auto-correlation emerge. At an observing frequency around 28 MHz ($z\sim 50$), a signal-to-noise of order 25 in the CMB cross-correlation would be achieved when the detection significance of the $21\,{\rm cm}$ auto-correlation would be only half as large, $\sim$12. Assuming white noise, and that foreground removal is possible, an experiment aiming to detect the $21\,{\rm cm}$ signal from the dark ages would be able to initially detect the CMB in cross-correlation long before the $21\,{\rm cm}$ signal emerges from the data. When the $21\,{\rm cm}$ auto-correlation reaches 0.5$\sigma$ the cross-correlation with the CMB would already by at 5$\sigma$.

\section{Instrumental Design and Observational Strategy Considerations}

In the previous section we established that a cross-correlation measurement, while challenging, is not implausible. Here, we consider more concrete instrumental considerations and perform a basic sensitivity calculation beyond a white noise treatment, highlighting some of the subtleties and contrasts with other low-frequency cosmology experiments.

In essence, the science goal of placing constraints on $x_{\rm HI} / T_S$ amounts to the measurement of the amplitude $\alpha$ of a template in
\begin{equation}
    T^{\rm radio} (\mathbf{\hat{n}}, \nu) = \alpha T^{\rm template} (\mathbf{\hat{n}}) + n(\mathbf{\hat{n}}, \nu),
\end{equation}
where $T^{\rm radio}$ is the observed low-frequency map, $T^{\rm template} (\mathbf{\hat{n}})$ is our template of CMB anisotropies, and $ n(\mathbf{\hat{n}})$ is our noise contribution. Alternatively, in harmonic space (convenient given the interferometric nature of many present and near-future instruments) we write
\begin{equation}
T^{\rm radio} (u, v, \mathbf{\hat{n}}) = \alpha T^{\rm template} (u, v, \mathbf{\hat{n}}) + n(u, v, \mathbf{\hat{n}}).
\end{equation}
Given the linearity of this measurement equation, it is straightforward to write down an optimal (minimum variance) estimator $\hat{\alpha}$ for the template amplitude:
\begin{equation}
\hat{\alpha} \equiv \frac{\sum_{u,v}\widetilde{T}^{\rm template,\dagger} (u,v) \sigma^{-2}(u,v) \widetilde{T}^{\rm radio} (u,v)}{\sum_{u,v} \sigma^{-2}(u,v) |\widetilde{T}^{\rm template} (u,v)|^2},
\end{equation}
where $\sigma(u,v)^2$ is the noise variance in a particular $uv$ mode, assuming a diagonal noise covariance in harmonic space (as is the case for interferometers, to a good approximation). The corresponding error $\Delta \hat{\alpha}$ on our estimator is then
\begin{equation}
\Delta \hat{\alpha} = \left[\sum_{u,v} \frac{|\widetilde{T}^{\rm template} (u,v)|^2}{\sigma^2 (u,v)} \right]^{-1/2}.
\end{equation}
To further simplify this, we assume radiometer noise (neglecting foreground power and intrinsic $21\,{\rm cm}$ fluctuations) for each baseline of an interferometer and obtain
\begin{equation}
\label{eq:alphaerr}
\Delta \hat{\alpha} = \frac{T_{\rm sys}}{T_{\rm sig}^{\rm eff}}\frac{1}{\sqrt{\Delta \nu \Delta t}},
\end{equation}
where
\begin{equation}
\label{eq:effectivesig}
    T_{\rm sig}^{\rm eff} \equiv \left(\sum_{u,v} C_{\ell = 2 \pi |\mathbf{u}|} N_{uv} \right)^{1/2}
\end{equation}
can be thought of as the effective signal as seen by an interferometer that has $N_{uv}$ copies of a baseline sampling a particular $uv$ mode, which probes the CMB fluctuations on an angular scale of $\ell = 2 \pi |\mathbf{u}|$ (adopting the flat-sky approximation for simplicity).

Equations~\ref{eq:alphaerr} and \ref{eq:effectivesig} provide a guide towards instrument design and observational strategy. One striking feature of the expression for $\Delta \hat{\alpha}$ is its simplicity, with no reference (for instance) to whether one's interferometer is operated as a tracking telescope or in a drift-scan mode. This is the result of our having a template of essentially negligible uncertainty. Crudely, the measurement of our template amplitude $\alpha$ involves the ratio of two maps, and as such it does not involve the breadth versus depth tradeoff to balance thermal noise and sample variance. The latter is not a source of error for a ratio measurement, and because $\alpha$ is a constant across the sky, thermal noise beats down at the same rate whether one integrates intently on a small patch or smears an observational campaign over a broad fraction of the sky.\footnote{This argument holds purely for one's sensitivity. In practice, ensuring the robustness of one's measurement would likely require surveying at least several different patches of sky.} Because of this insensitivity to survey strategy, a measurement of $\hat{\alpha}$ may be an attractive commensal target for low-frequency observations.

    \begin{figure}[!t]
        \centering
    	\includegraphics[width=1.0\textwidth]{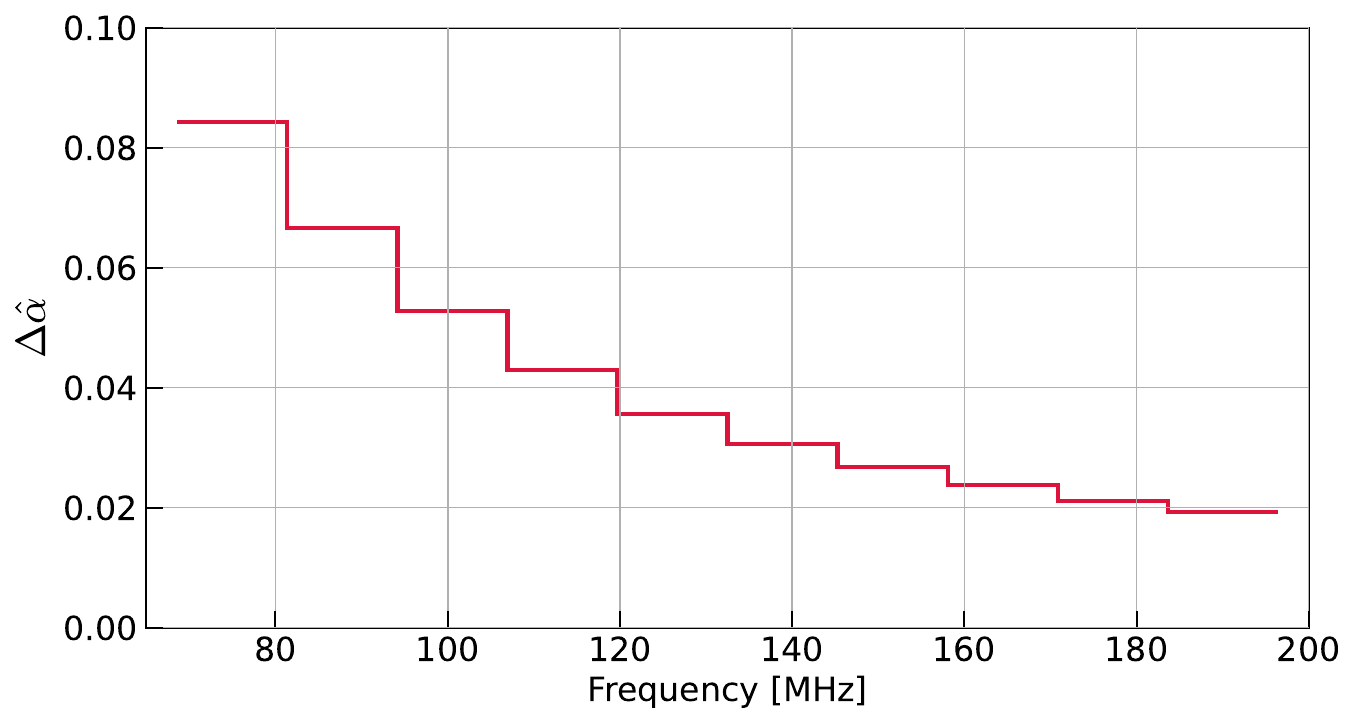}
    	\caption{Forecasted errors as a function of frequency for a template-based amplitude measurement, assuming a $30 \times 30$ interferometric array consisting of $6\,\textrm{m}$ dishes, observing for $\Delta t = 4000\,\textrm{hours}$, shown in bins of 12.8 MHz. The resulting sensitivities are within a reasonable range for detecting $\tau(\nu)$ effects.}
        \vspace{0.5cm}
        \label{fig:deltaalphaerr}
    \end{figure}

In contrast to observational strategy, instrumental design strongly affects one's sensitivity to $\alpha$. Equation~\ref{eq:effectivesig} suggests deploying an interferometer with baselines $\mathbf{b}$ that probe $uv$ modes $\mathbf{u} = \mathbf{b} / \lambda$ where CMB anisotropies are strongest. Since $C_{\ell = 2 \pi |\mathbf{u}|}$ is a decreasing function of $\ell$, this pushes one towards compact arrays with short baselines. Assuming a sky-noise dominated system temperature modeled using the Global Sky Model \citep{2008MNRAS.388..247D, 2017MNRAS.464.3486Z}, one arrives at\footnote{The exact expression is not precisely a product of power laws, but Equation~\eqref{eq:alphahaterr} is a reasonable approximation across the relevant parameter space. For our fiducial array it is good to sub-percent levels at the high-frequency end and $\sim 10\%$ at the low-frequency end. For very small arrays the fit degrades to factor of $\sim 2$ accuracy, but this is largely irrelevant anyway since such arrays do not possess sufficient sensitivity for our proposed measurements.}
\begin{equation}
\label{eq:alphahaterr}
    \Delta \hat{\alpha}(\nu) = 0.665 \left(\frac{\nu}{ 100\,\textrm{MHz}}\right)^{-1.75}  \left( \frac{\Delta \nu}{0.1\,\textrm{MHz}} \right)^{-1/2} \left( \frac{\Delta t}{4000\,\textrm{hr}} \right)^{-1/2}   \left( \frac{n_{\rm sq}}{30}\right)^{-1.08} \left(\frac{D}{6\,\textrm{m}}\right)^{0.91},
\end{equation}
where $\nu$ is the observing frequency, $\Delta \nu$ is the channel width, $D$ is the dish diameter, and we have assumed a square array of $n_{\rm sq} \times n_{\rm sq}$ close-packed dishes (i.e., the shortest baselines are of length $D$). One sees that smaller dishes are weakly preferred, mostly because one cannot have shorter baselines than the physical size of a dish. As expected, constructing more dishes provides more baselines for increased sensitivity. The frequency dependence of $\Delta \hat{\alpha}(\nu)$ is the product of two competing factors: on one hand, the sky is brighter at low frequencies, which increases $T_{\rm sys}$ and thus the noise; on the other hand, for a baseline of fixed physical length, lower frequency observations probe coarser angular scales, where the CMB fluctuations are more prominent, increasing the signal. The former effect is the dominant one, but the latter moderates the overall frequency dependence such that the dependence on $\nu$ is reduced from $\nu^{-2.7}$ (the sky model's prediction) to $\nu^{-1.75}$. 

In Figure~\ref{fig:deltaalphaerr} we show $\Delta \hat{\alpha}(\nu)$ for a hypothetical survey with $30 \times 30$ array of $6\,\textrm{m}$ dishes, integrating for $\delta t = 4000\,\textrm{hours}$ from $75\,\textrm{MHz}$ to $200\,\textrm{MHz}$. The results are rebinned into 10 coarse bins. Since the sensitivity scales as $1/ \sqrt{\Delta \nu}$, there is no preference for a fine spectral resolution survey that is rebinned after the fact over coarser channels widths, at least as far as sensitivity is concerned. This amounts to an overall significance $\ga 100\sigma$ overall, and one can see that it is in the ballpark for detecting percent-level effects induced by $\tau_{\rm HI}$ seen in Figure~\ref{fig:tau}.

Our sensitivity calculation here neglects foregrounds, which will likely severely impact the feasibility of measuring $\alpha(\nu)$. The primary challenge is that even though $\tau_{\rm HI}$ does have some interesting features, it is still a relatively smooth function (like the foregrounds; \citealt{2020PASP..132f2001L}) but is a small-amplitude effect (unlike the foregrounds). Filtering out the smoothest component to fight foregrounds can significantly reduce the amplitude of $\tau_{\rm HI}$, necessitating increased signal-to-noise and mitigation strategies that are as non-aggressive as possible. To this end, new methods for low-frequency foreground removal that do not rely on spectral smoothness (Liu et al., in preparation; Fronenberg \& Liu, in prepration) and array designs that minimize spectral leakage \citep{2018ApJ...869...25M,2026ApJ...999...96M} by reducing the ``foreground wedge" are desirable. 
Other promising techniques may include blind and semi-blind component separation methods that exploit the statistical and morphological differences between diffuse foregrounds and cosmological fluctuations, such as Independent Component Analysis (ICA; \cite{2012MNRAS.423.2518C}) and Generalized Morphological Component Analysis (GMCA; \cite{2013MNRAS.429..165C}). Gaussian-process (GP) foreground modeling provides an additional probabilistic framework in which foregrounds and the CMB signals are described by distinct covariance kernels, enabling flexible modeling of correlated spectral structure while naturally propagating uncertainties (e.g., \cite{2018MNRAS.478.3640M}). Another promising direction is joint Bayesian forward modeling, in which foregrounds, instrumental response, calibration uncertainties, and the target cosmological signal are simultaneously modeled within a unified framework. Such approaches can incorporate both angular and spectral information, reduce the risk of signal loss from aggressive foreground filtering, and are increasingly being explored through differentiable end-to-end inference frameworks such as BayesLIM \citep{2025MNRAS.541..687K}.

Put together, an attempt to measure $\alpha(\nu)$ may be an attractive target for upcoming low-frequency arrays. Its survey strategy agnosticism makes it ideal for commensal observations that take place while other science observations are pursued. The desire for a compact array will test mutual coupling mitigation requirements for next-generation receiving elements, and the pursuit of minimally invasive foreground cleaning strategies will push new wedge reduction array design and methods that do not rely on spectral smoothness. At the same time, $\tau_{\rm HI}$ measurements probe much of the same physics as global $21\,\textrm{cm}$ experiments do, but have the advantage of being a mapping experiment that has a very well-measured template to pursue and more varied opportunities for jackknives and null tests. In this way, interferometers may be able to more directly weigh in on the EDGES anomaly \citep{2018Natur.555...67B} than the usual proposals of jointly modeling the $21\,\textrm{cm}$ global signal and $21\,\textrm{cm}$ fluctuations with uncertain astrophysical prescriptions. 

\section{Conclusion}
\label{sec:concl}

We have shown that CMB fluctuations measured at radio wavelengths have an overall frequency-dependent amplitude coming from absorption by neutral hydrogen at high redshift. This frequency dependence is sensitive to the neutral fraction $x_{\rm HI}$ at the redshift that maps $\nu$ into the rest frequency of the $21\,{\rm cm}$ transition and to the spin temperature $T_s$ of the neutral hydrogen at that redshift in the combination $x_{\rm HI}/T_s$. By cross-correlating radio measurements of the CMB with mm-wave measurements it should be possible to measure this effect. This opens up a new view on $21\,{\rm cm}$ cosmology.

By measuring the primary CMB fluctuations in cross-correlation, it is possible that contamination by Galactic and extragalactic foregrounds will present problems in different ways from those encountered by experiments aiming to detect the fluctuations in the $21\,{\rm cm}$ emission or the evolution of the mean brightness temperature. In terms of raw signal-to-noise, we find that the detection of the primary CMB in cross-correlation does not require prohibitively low noise levels. In fact, for experiments targeting the ``dark ages'' we find that it may be easier to detect the CMB in cross-correlation than it will be to detect the fluctuations in the neutral hydrogen. The CMB fluctuations in cross-correlation could be a useful intermediate target for these experiments, even if the small power reduction from the excess radio optical depth is too difficult.

As is the case with the intrinsic $21\,{\rm cm}$ fluctuations, the difficulty is likely to come from separating out systematic effects and foreground contamination. An optimized strategy for CMB detection at radio wavelengths may look very different than one searching for $21\,{\rm cm}$ fluctuations, and thus we leave calculations, end-to-end simulations (e.g., in the style of \citealt{2024ApJ...975..222F}), and detailed further speculation for future work.

\section*{Acknowledgments}
We thank Olivier Dore, Hannah Fronenberg, Adam Lidz, Mat Madhavacheril, and Jon Sievers for useful discussions.  GH thanks Brand and Monica Fortner for support. AL acknowledges support from the Natural
Sciences and Engineering Research Council of Canada
through their Discovery Grants Program and their Alliance International Program, as well as
the William Dawson Scholar program at McGill University. Part of the research was carried out at the Jet Propulsion Laboratory (JPL), California Institute of Technology, under a contract with the National Aeronautics and Space Administration (80NM0018D0004). DZ was partially supported by a Grainger Distinguished Postdoctoral Fellowship. This research was supported in part by grant no. NSF PHY-2309135 to the Kavli Institute for Theoretical Physics (KITP).

\bibliographystyle{aasjournal}

\bibliography{radio_fog}

\end{document}